\documentstyle[cite,sprocl,epsfig]{article} 
 
\newcommand{\andvol}[3]{{\bf #1}~#2~(#3)}
\newcommand{\PRL}[3]{Phys.~Rev.~Lett.~\andvol{#1}{#2}{#3}}
\newcommand{\PRD}[3]{Phys.~Rev.~\andvol{D#1}{#2}{#3}}

\newcommand{\NPB}[3]{Nucl.~Phys.~\andvol{B#1}{#2}{#3}}
\newcommand{\PLB}[3]{Phys.~Lett.~\andvol{B#1}{#2}{#3}}

\renewcommand{\Im}{{\rm Im}}
\newcommand{\SACD}{S_{\rm ACD}}
\newcommand{\Del}[1]{\Delta_{\rm #1}}
\newcommand{\polys}{p_N(s)}
\newcommand{\beq}{\begin{equation}}
\newcommand{\eeq}{\end{equation}}
\newcommand{\beqx}{\begin{displaymath}}
\newcommand{\eeqx}{\end{displaymath}}
\newcommand{\beqa}{\begin{eqnarray}}
\newcommand{\eeqa}{\end{eqnarray}}
\newcommand{\beqax}{\begin{eqnarray*}}
\newcommand{\eeqax}{\end{eqnarray*}}
\begin{document}

\begin{titlepage}

\begin{flushright}
UCTP--2/97 \\
VPI--IPPAP--97--2 \\
CERN--TH/96--198 \\
hep--ph/9702439
\end{flushright}

\bigskip
\vfill

\begin{center}
{\large\bf 
 AN EVALUATION OF \\ THE ANALYTIC CONTINUATION BY DUALITY TECHNIQUE
}\\
\bigskip
\bigskip
{\normalsize
 {\sc  T.~Takeuchi
 } \\
 \smallskip
 {\it Institute for Particle Physics and Astrophysics, \\ 
      Physics Department, Virginia Tech, Blacksburg, VA 24061-0435\ \footnote{
  Current address.}
 } \\
 \medskip 
 {\sc L.~C.~Goonetileke, S.~R.~Ignjatovi\'c, and L.~C.~R.~Wijewardhana
 } \\
 \smallskip
 {\it Department of Physics, University of Cincinnati, Cincinnati, OH 45221
 } \\
}
\end{center}

\bigskip
\bigskip

\begin{abstract}
In Ref.~\citen{SUND:93}, the value of the
oblique correction parameter $S$ for walking technicolor
theories was estimated using a technique called 
{\it Analytic Continuation by Duality} (ACD). 
We apply the ACD technique to the perturbative vacuum polarization 
function and find that it fails to reproduce the well known
result $S=1/6\pi$. 
This brings into question the reliability of the ACD technique and
the ACD estimate of $S$.
\end{abstract}

\bigskip
\bigskip

\begin{center}
{\it Talk presented at the 1996 International Workshop on \\ 
     Perspectives of Strong Coupling Gauge Theories (SCGT'96) \\
     13--16 November, 1996, Nagoya, Japan.}
\end{center}

\bigskip
\bigskip

\vfill

\begin{flushleft}
UCTP--2/96 \\
VPI--IPPAP--97--2 \\
CERN--TH/96--198 \\
hep--ph/9702439 \\
February 1997
\end{flushleft}

\end{titlepage}

\title{AN EVALUATION OF \\ THE ANALYTIC CONTINUATION BY DUALITY TECHNIQUE}
 
\author{T.~TAKEUCHI\ \footnote{Presenting author.}}
\address{TH Division, CERN, CH--1211 Gen\`eve 23, Switzerland}
\author{L.~C.~GOONETILEKE, S.~R.~IGNJATOVI\'C, and L.~C.~R.~WIJEWARDHANA}
\address{Department of Physics, University of Cincinnati, Cincinnati, OH 45221}

\maketitle

\abstracts{
In Ref.~\citen{SUND:93}, the value of the
oblique correction parameter $S$ for walking technicolor
theories was estimated using a technique called 
{\it Analytic Continuation by Duality} (ACD). 
We apply the ACD technique to the perturbative vacuum polarization 
function and find that it fails to reproduce the well known
result $S=1/6\pi$. 
This brings into question the reliability of the ACD technique and
the ACD estimate of $S$.
}

%%%%%%%%%%%%%%%%%%%%%%%%%%%%%%%%%%%%%%%%%%%%%%%%%%%%%%%%%%%%%%%%%%%%%%%%%%%%%%%
%%%%%%%%%%%%%%%%%%%%%%%%%%%%%%%%%%%%%%%%%%%%%%%%%%%%%%%%%%%%%%%%%%%%%%%%%%%%%%%

\section{Introduction}

The {\it analytic continuation by duality} (ACD) technique
was proposed in Ref.~\citen{SUND:93} 
as a potentially reliable method to compute the oblique correction parameter 
$S$ for technicolor theories.
The advantage of the ACD technique was that it could be
applied to both QCD--like and walking technicolor \cite{APPE:86}
theories whereas the dispersion relation technique used by 
Peskin and one of us
in Ref.~\citen{PESK:90} could only be applied to the former.
Furthermore, the ACD estimate of $S$ for walking technicolor
implied that walking dynamics could render $S$ negative,
making it compatible with the current experimental limit. \cite{HEWE:96}
This was in contrast to the result of Harada and Yoshida \cite{HARA:94}
who used the Bethe--Salpeter equation approach to conclude that
$S$ was positive even for walking theories.

In this talk, we investigate the reliability of the ACD technique.
In section 2, we review the definition of the $S$ parameter and 
explain the ACD technique.
In section 3, we apply the ACD technique to the perturbative spectral
function to see if the famous result $1/6\pi$ could be reproduced.
Discussions and conclusions are stated in Section 4.

\section{The ACD technique}

\begin{figure}[t]
\begin{center}
\epsfig{file=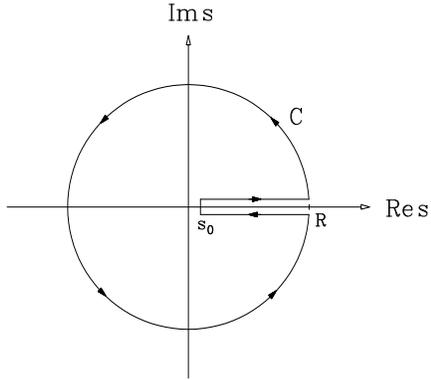,angle=90,height=5cm,width=5.6cm}
\caption{The contour $C$ which avoids the branch cut along the
           real $s$--axis.}
\label{CONTOUR}
\end{center}
\end{figure}

The $S$ parameter, as defined in Ref.~\citen{PESK:90}, is equal to
a certain linear combination of electroweak vacuum polarization
functions evaluated at zero momentum transfer.
We represent this schematically as
\beqx
S = \Pi(0).
\eeqx
(The precise definition of $\Pi(s)$ is irrelevant to our
ensuing discussion.)
The vacuum polarization function $\Pi(s)$ is analytic in the
entire complex $s$ plane except for a branch cut along the
positive real $s$ axis starting from the lowest particle threshold 
contributing to $\Pi(s)$.   
Applying Cauchy's theorem to the contour 
$C$ shown in Fig.~\ref{CONTOUR}, we find
\beq
S = \frac{1}{\pi}\int_{s_0}^R ds \frac{ \Im\Pi(s) }{ s }
  + \frac{1}{2\pi i}\oint_{|s|=R} ds \frac{ \Pi(s) }{ s }.
\label{CAUCHY1}
\eeq
If the radius of the contour $R$ is taken to infinity, the integral
around the circle at $|s|=R$ can be shown to vanish 
and we obtain the dispersion relation
\beqx
S = \frac{1}{\pi}\int_{s_0}^R ds \frac{ \Im\Pi(s) }{ s },
\eeqx
which was used in Ref.~\citen{PESK:90} to calculate $S$.
However, the dispersion relation approach requires the knowledge of
$\Im\Pi(s)$ along the real $s$ axis
which is only available for QCD--like technicolor theories.

The basic idea of the ACD technique, on the other hand, 
is to approximate the kernel $1/s$ by a polynomial
\beqx
\frac{1}{s} \approx \polys = \sum_{n=0}^{N} a_n(N) s^n,
\qquad s\in[s_0,R]
%\label{FIT}
\eeqx
and use it to make the integral along the real $s$ axis vanish instead.
Applying Cauchy's theorem to the product $p_N(s)\Pi(s)$ over the 
same contour $C$ yields 
\beq
0 = \frac{1}{\pi}\int_{s_0}^R \!ds\,\polys\Im\Pi(s)
  + \frac{1}{2\pi i}\oint_{|s|=R}\!ds\,\polys\Pi(s).
\label{ZERO}
\eeq
Subtracting Eq.~(\ref{ZERO}) from Eq.~(\ref{CAUCHY1}), we obtain
\beqx
S = S_N + \Del{fit},
\eeqx
where
\beqax
S_N & \equiv & \frac{1}{2\pi i}\oint_{|s|=R} \!ds
              \left[ \frac{1}{s} - \polys \right] \Pi(s), \cr
\Del{fit}
   & \equiv & \frac{1}{\pi}\int_{s_0}^R \!ds 
              \left[ \frac{1}{s} - \polys \right] \Im\Pi(s).
\eeqax
For sufficiently large $N$, we can expect $\Del{fit}$ to
be negligibly small.  In fact, it converges to zero in the
limit $N\rightarrow \infty$ (though how quickly the convergence
occurs depends on the interval $[s_0,R]$).
We can therefore neglect it and approximate
$S$ with $S_N$ which is an integral around the circle
$|s|=R$ only.   We call $\Del{fit}$ the {\it fit error}.

If the radius of the contour $R$ is taken to be sufficiently large,
the function $\Pi(s)$ can be approximated on $|s|=R$ 
by a large momentum expansion: 
\beq
\Pi(s) \approx \sum\limits_{m=1}^M \frac{b_m(s)}{s^m}.
\label{ope}
\eeq
This expression is obtained by analytically continuing the 
operator product expansion (OPE) of $\Pi(s)$ from the deep Euclidean
region where it can be calculated for both QCD--like and
walking technicolor theories.
Therefore, we can write
\beqx
S_N = S_{N,M} + \Del{tr},
\eeqx
where
\beqax
S_{N,M}  & \equiv &  \frac{1}{2\pi i}\oint_{|s|=R} \!ds
                     \left[ \frac{1}{s} - \polys
                     \right]
                     \sum_{m=1}^M \frac{b_m(s)}{s^m}, \cr
\Del{tr} & \equiv &  \frac{1}{2\pi i}\oint_{|s|=R} \!ds
                     \left[ \frac{1}{s} - \polys
                     \right] 
                     \left[ \Pi(s)
                          - \sum_{m=1}^M \frac{b_m(s)}{s^m}
                     \right],
%\label{trunc}
\eeqax
and approximate $S_N$ with $S_{N,M}$.  The neglected term
$\Del{tr}$ is called the {\it truncation error}.

It is often the case that the approximation is taken one step further
by neglecting the $s$--dependence of the expansion 
coefficients in Eq.~(\ref{ope}),
{\it i.e.}
\beqx
b_m(s) \approx b_m(-R) \equiv \hat{b}_m.
\eeqx
This is obviously a dangerous approximation to make since the
analytic structure of the integrand will be completely altered.
Define
\beqx
S_{N,M} = \SACD + \Del{AC}
\eeqx
where
\beqax
\SACD & \equiv & \frac{1}{2\pi i}\oint_{|s|=R}\!ds
               \left[ \frac{1}{s} - \polys
               \right] \sum_{m=1}^M \frac{\hat{b}_m}{s^m}, \cr
\Del{AC} & \equiv & \frac{1}{2\pi i}\oint_{|s|=R}\!ds
               \left[ \frac{1}{s} - \polys
               \right] \sum_{m=1}^M \frac{b_m(s)-\hat{b}_m}{s^M}.
\eeqax
It can be argued that $\Del{AC}$ is highly
suppressed and thus negligible since the difference 
$1/s-\polys$ is approximately zero in the vicinity of the
positive real $s$ axis where the difference $b_m(s)-\hat{b}_m$
can be expected to be most pronounced. Thus:
\beqx
S \approx \SACD.
\eeqx
In this approximation, the integral for $\SACD$ will only pick
up the residues of the single poles inside the integration contour
and we find,
\beqx
\SACD = -\sum\limits_{n=0}^{\min\{N,M-1\}}a_n(N) \hat{b}_{n+1}.
%\label{Sacd}
\eeqx
We will call $\Del{AC}$ the {\it analytical continuation error}. 

To summarize, the ACD technique uses the relation
\beqx
S = \SACD + \Del{AC} + \Del{tr} + \Del{fit},
\eeqx
and assumes that all three types of error can be neglected and
approximates $S$ with $\SACD$.

\section{The Perturbative Spectral Function}

\begin{table}[t]
\caption{$\SACD$ and the fit, truncation, and analytical continuation
errors for the perturbative vacuum polarization function. 
The cutoffs are $[s_0,R] = [4m^2,25m^2]$, and the fit routine was the
least square fit.  The exact value of $S$ is $1/6\pi = 0.0531$. 
\hspace*{\fill}}
\label{tab1}
\begin{center}
\begin{tabular}{cccccc}
\hline
\hline
$N$ & $M$ & $\SACD$ & $S_{N,M}=\SACD+\Del{AC}$ & $\Del{fit}$ & $\Del{tr}$
 \\
\hline\hline
$3$ & $2$ &  $0.2930$ &  $0.0580$ & $-0.0002$ & $-0.0048$  \\
    & $3$ &  $0.2883$ &  $0.0530$ &           & $\phantom{-}0.0002$ \\
    & $4$ &  $0.2884$ &  $0.0532$ &           & $-0.0000$  \\
\hline
$4$ & $2$ &  $0.4330$ &  $0.0632$ & $-0.0001$ & $-0.0101$  \\
    & $3$ &  $0.4203$ &  $0.0521$ &           & $-0.0010$  \\
    & $4$ &  $0.4211$ &  $0.0532$ &           & $-0.0000$  \\
    & $5$ &  $0.4211$ &  $0.0531$ &           & $\phantom{-}0.0000$ \\
\hline
$5$ & $3$ &  $0.5731$ &  $0.0506$ & $-0.0000$ & $\phantom{-}0.0025$ \\
    & $4$ &  $0.5759$ &  $0.0533$ &           & $-0.0002$  \\
    & $5$ &  $0.5757$ &  $0.0531$ &           & $\phantom{-}0.0000$ \\
    & $6$ &  $0.5757$ &  $0.0531$ &           & $-0.0000$  \\
\hline\hline
\end{tabular}
\end{center}
\end{table}

To check validity of the approximation $S\approx\SACD$,
we calculate $\SACD$ for the one--loop contribution of a massive
fermion doublet to $S$.
The vacuum polarization function $\Pi(s)$ in this case is
given by:
\beq
\Pi_{\rm pert}(s) =
-\frac{1}{\pi}\frac{m^2}{s}
\int_0^1 \!dx\log\left[ 1 - x(1-x)\frac{s}{m^2}
                 \right].
\label{FPERT}
\eeq
Evaluating this expression at $s=0$, we find the well known result
$S=1/6\pi$.

The function $\Pi_{\rm pert}(s)$ is analytic in the entire complex $s$ plane 
except for a branch cut along the positive real $s$ 
axis starting from $s=4m^2$.
The imaginary part of this function along the cut is given by
\beq
\Im\Pi_{\rm pert}(s) = \frac{m^2}{s}\beta \theta(s-4m^2),
\qquad
\beta = \sqrt{1-\frac{4m^2}{s}}.
\label{IMFPERT}
\eeq
The first few terms of the large $s$-expansion of $\Pi_{\rm pert}(s)$
are given by
\beqax
\lefteqn{\pi\Pi_{\rm pert}(s)} \cr
& = & x
      \left\{ -\ln\left(-\frac{1}{x}\right)+2
      \right\}
    + x^2 
      \left\{ 2\,\ln\left(-\frac{1}{x}\right)+2
      \right\}
    + x^3 
      \left\{ 2\,\ln\left(-\frac{1}{x}\right)-1
      \right\}
+ \dots,
\eeqax
where $x\equiv 4m^2/s$.
Using these expressions, we calculated $\SACD$,
$\Del{AC}$, $\Del{tr}$, and $\Del{fit}$.
The results of our calculations for several values of $N$ and $M$
are shown in Table~\ref{tab1}.
The fit interval was $[s_0,R]=[4m^2,25m^2]$,
and the fit routine was the least square fit.

As is evident from Table.~\ref{tab1},
the fit and truncation errors are under excellent control and
$S_{N,M}$ reproduces the exact value of $S$ accurately
already at $N=M=3$.
However, the analytic continuation error is not.
For the $N=5$ case, for instance,  $\SACD$ is larger 
than the exact value by more than an order of magnitude.
In fact, we find that $\SACD$ and $\Del{AC}$ diverge 
as $N\rightarrow\infty$.   

We conclude that neglecting the $s$--dependence
of the $b_m(s)$'s fails miserably as an approximation.
The reason for this can be traced to the fact that even 
though the difference $1/s - \polys$ converges to zero within
its radius of convergence, outside it diverges.
Therefore, the handwaving argument of the previous section was wrong:
the error induced by the neglect of the $s$--dependence of the
$b_m(s)$'s may be suppressed near the real $s$ axis, but it is actually
{\it enhanced} away from it.

\section{Discussion and Conclusions}

The application of the ACD technique to the perturbative vacuum
polarization function has shown that the analytic continuation error
$\Del{AC}$ is not under control and that the approximation $S\approx\SACD$
cannot be trusted.  
This brings into doubt the reliability of the
ACD estimate of $S$ obtained in Ref.~\citen{SUND:93}.

A natural question to ask next is whether the ACD technique can be improved
by including the $s$--dependence of the large momentum expansion coefficients
$b_m(s)$ and using $S_{N,M}$ as the estimate of $S$ instead of $\SACD$.
In the perturbative case, we have seen that this is an excellent 
approximation.    However, whether $S_{N,M}$ will reproduce the correct
value of $S$ for all cases is far from clear.   If the large momentum
expansion is an asymptotic series, the truncation error $\Del{tr}$ may
not converge to zero in the limit $M\rightarrow\infty$.
Even if it is a convergent series, the convergence may be too slow
for the method to be practical.
In a toy model with a spectral function $\Im\Pi(s)$ which
is representative of the QCD spectrum, we have found that the inclusion
of the $s$--dependence in the large momentum expansion does not
necessarily improve the estimation of $S$.
This, and other related problems will be 
discussed in subsequent papers. \cite{next}

%%%%%%%%%%%%%%%%%%%%%%%%%%%%%%%%%%%%%%%%%%%%%%%%%%%%%%%%%%%%%%%%%%%%%%%%%%%%%%%
\newpage

\section*{Acknowledgments} 

We would like to thank M.~E.~Peskin, R.~Sundrum,
and K.~Takeuchi for helpful discussions.
This work was supported in part by the United States Department of Energy 
under Contract Number DE--AC02--76CH03000 (T.T.) 
and Grant Number DE--FG02--84ER40153 (L.C.G., S.R.I., and L.C.R.W.).

\end{document}